\documentclass[twocolumn,preprintnumbers,amsmath,amssymb]{revtex4}
\pdfoutput=1 
\usepackage{dcolumn}
\usepackage{color}
\usepackage{bm}
\usepackage{graphicx}
\usepackage{epsfig}
\usepackage{amsmath}
\usepackage{amsfonts}
\usepackage{amssymb}

\def \e{\mathrm{e}}
\def \half{{1 \over 2}}
\def \pxipy{$p_x\!+\!ip_y$ }
\def \He3A{$^3$HeA}
\def \SRO{{Sr$_2$RuO$_4$}}
\def \omgvec{\boldsymbol{\Omega}}
\def \D02ef{\Delta_{0\phantom{2}}^{\phantom{0}2}/\epsilon_F}
\def \etal{\textit{et al.}~}

\begin{document}
\draft

\title{
    Testing for Majorana Zero Modes in a \pxipy Superconductor at High Temperature by Tunneling Spectroscopy}
\author{
    Yaacov E.~Kraus$^{1}$, Assa Auerbach$^{1}$,
    H.A.~Fertig$^{2}$ and Steven H.~Simon$^{3}$ }
\affiliation{
    1) Physics Department, Technion, 32000 Haifa, Israel\\
    2) Department of Physics, Indiana University, Bloomington, IN 47405, USA\\
    3) Alcatel-Lucent, Bell Labs, 600 Mountain Avenue, Murray Hill, NJ 07974, USA}

\begin{abstract}
Directly observing a zero energy Majorana state in the vortex core
of a chiral superconductor by tunneling spectroscopy requires energy
resolution better than the spacing between core states $\D02ef$. We
show that nevertheless, its existence can be decisively detected by
comparing the temperature broadened tunneling conductance of a
vortex with that of an antivortex even at temperatures $T \gg\
\D02ef$.
\end{abstract}
\pacs{74.50.+r ,67.50.Fi,03.67.Lx, 71.10.Pm, 74.20.Rp} %
\maketitle


Driven partially by the dream of building naturally error
resistant quantum computers, the study of topological phases of
matter has become an important topic of research
\cite{NayakRMP}. The simplest class of topological phases of
matter that could be useful in this respect are the chiral \pxipy
BCS paired systems \cite{Bravyi}. There are several physical
systems where \pxipy pairing is believed to be realized, including
the A phase of superfluid $^3$He \cite{He3Volovik} (\He3A), the exotic
superconductor \SRO \cite{RiceSigrist}, and the $\nu=5/2$ quantum
Hall state \cite{MooreRead,Grieter}.  In addition there have been
recent proposals to realize  \pxipy pairing in cold fermion gases
\cite{GurarieRadzihosky}. In  these (weak) \pxipy systems, certain
types of vortices (quasiparticles in the quantum Hall context
\cite{ReadGreen}) are believed to carry zero energy Majorana
fermions \cite{Volovik,ReadGreen} which are the topologically
protected degrees of freedom.

In \SRO $\, $ and \He3A the vortices that carry the Majorana
fermions are the so-called half-quantum vortices, which can be
thought of as a vortex in the order parameter of one spin species
without a vortex in the order parameter of the opposite species
\cite{TewariKim}. (Note that in spin-polarized \pxipy systems,
including proposed atomic gas realizations or the 5/2 state, there
is no half-quantum vortex and the  full quantum vortex carries the
Majorana fermions.)

Let us suppose that in one of these systems, the relevant
Majorana-fermion-carrying vortex has been observed
\cite{e4quantumHall}. The next important step would be to design an
experiment to observe the Majorana fermion in such a vortex
\cite{DemlerTewari}. In the case of \SRO, one obvious experiment
would be an energy-resolved tunneling experiment, which measures the
local density of states (LDOS) \cite{Shore}. An observation of a
localized mode at precisely zero energy would be direct evidence of
the Majorana mode. For cold atoms, an analogous experiment for
observing the LDOS would be an energy-resolved local particle
annihilation experiment. For the other realizations of \pxipy order
it is not as clear how such an experiment would be performed
\cite{endnote1}.

\begin{figure}[htb]
\vspace{-0.3cm}
\begin{center}
\includegraphics[width=8cm,angle=0]{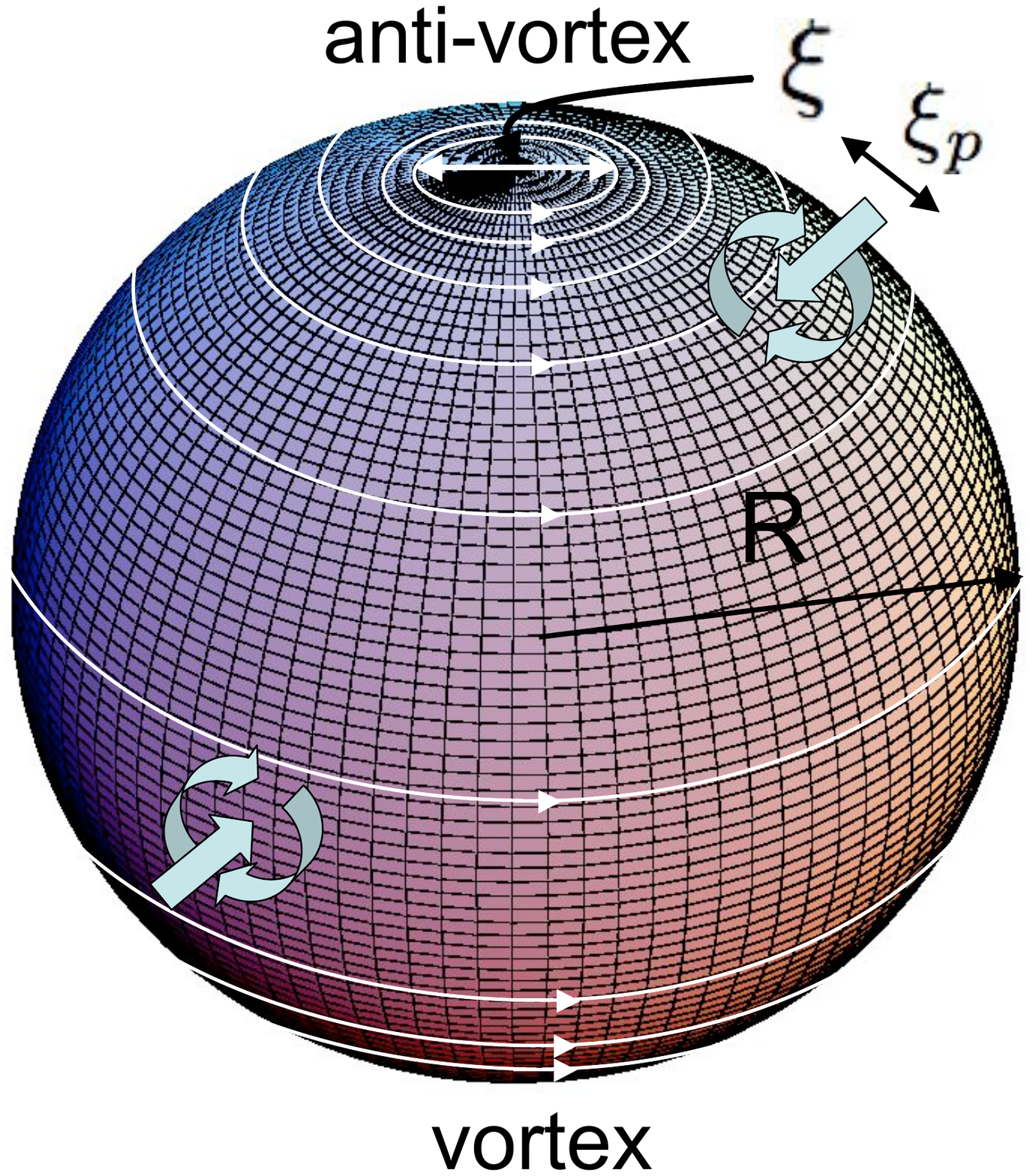}
\vspace{-0.8cm} %
\caption{  \label{Fig:sphere} %
A vortex pair of the \pxipy superconductor on the sphere, described
by Eq.~\ref{Eq:Delta_vv}. Thin white lines represent the current
flow. Wide arrows represent the pair relative angular momentum.
$\xi_p$ is the pairing range. $\xi$ is the coherence length which
determines the vortex core size. }
\end{center}
\end{figure}

In principle such tunneling experiments could provide definitive
evidence for the Majorana mode. However, in practice they may be
prohibitively difficult. In the vortex, there will exist sub-gap
bound states in the core known as Caroli-de-Gennes-Matricon (CdGM)
states \cite{CdGM,KopninSalomaa}. The spacing between these bound
states is typically of order $\delta_c = \D02ef$ where $\Delta_0$ is
the gap (presumably on order of the critical temperature) and
$\epsilon_F$ is the Fermi energy. Since the experimentally observed
tunneling spectrum will be smeared by the temperature, this
tunneling experiment would naively only have a clear signature for
$T < \delta_c$. Unfortunately such low temperatures could
potentially be unattainable in any of the proposed realizations
($\delta_c \approx 7 \mu K$ in \He3A, $ <  0.1 mK$ in \SRO). The
purpose of this paper is to demonstrate that the tunneling spectrum
retains an unambiguous signature of the Majorana fermion at much
higher temperatures. The signature is found by comparing the
tunneling conductance peaks of a vortex with an antivortex, the
direction of vorticity being defined relative to the angular
momentum of the chiral order parameter.


\begin{figure}[htb]
\vspace{-0.3cm}
\begin{center}
\includegraphics[width=9cm,angle=0]{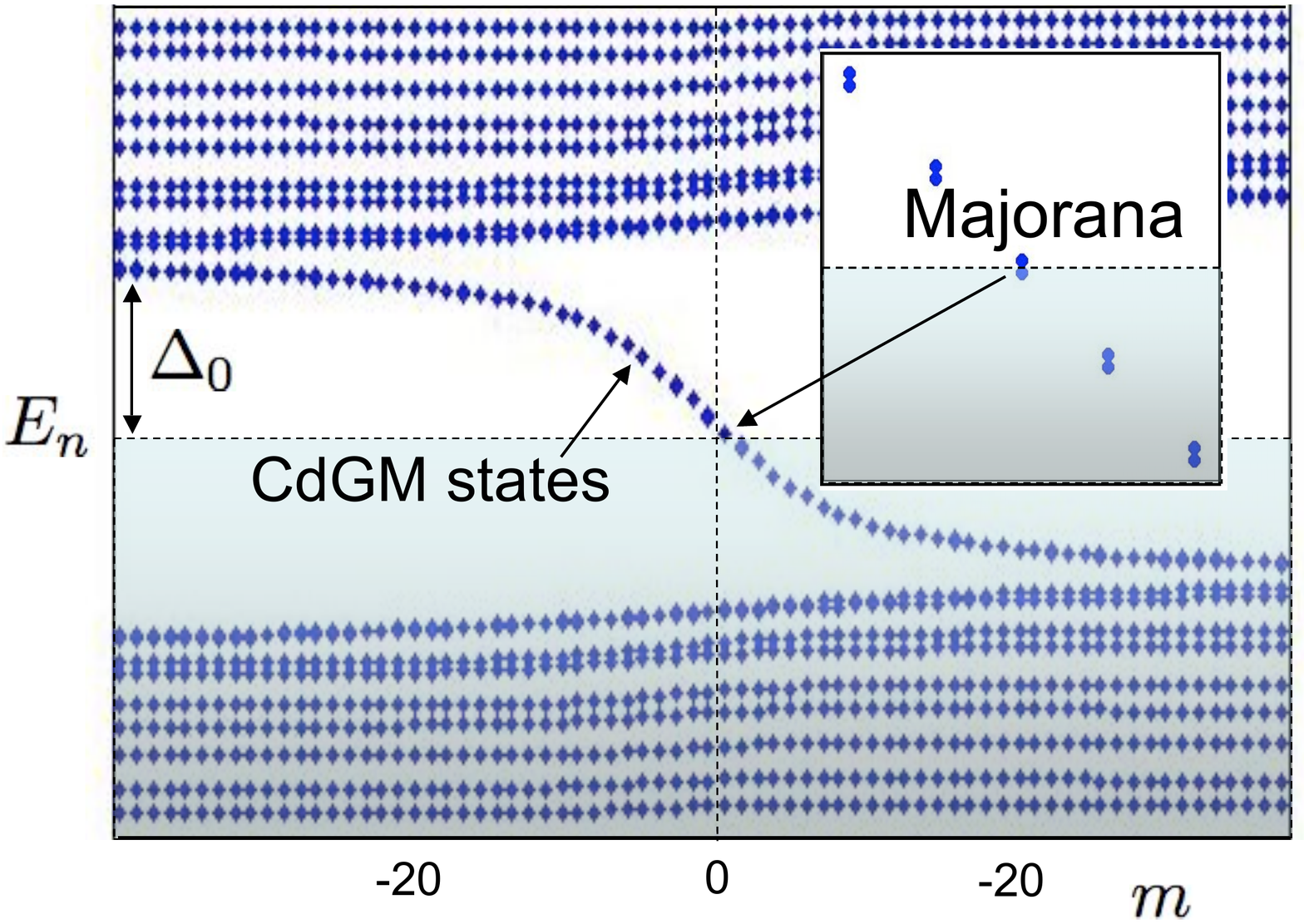}
\vspace{-0.8cm} %
\caption{ \label{Fig:Em} %
BdG spectrum $E_{n(m)}$, of the vortex pair on the sphere, depicting
the CdGM core states. The inset shows that their  double
degeneracies are split by weak tunneling between the poles. The
state nearest zero energy is the Majorana mode of both vortex and
antivortex. }
\end{center}
\end{figure}

{\em Bogoliubov de-Gennes (BdG) theory. }
We consider a two dimensional uniform \pxipy superconductor of
spinless fermions. The BdG excitations are given by \cite{BdG}
\begin{equation} \label{Eq:BdG}
    \left( \begin{array}{cc}
           \widehat{T}-\epsilon_F      & \Delta \\
           \Delta^{\dagger}            & -(\widehat{T}-\epsilon_F) \end{array} \right)
    \left( \begin{array}{c} u_n \\ v_n \end{array} \right)
    = E_n \left( \begin{array}{c} u_n \\ v_n \end{array} \right),
\end{equation}
where $\widehat{T}$ is the kinetic energy operator, and $\epsilon_F$
is the Fermi energy.

We implement the BdG equation on a sphere of radius $R$,
parameterized by the unit vector $\omgvec = (\theta,\phi)$. The
spherical geometry has two important advantages: (i) It has no
boundaries, which strongly affect the low energy spectrum. (ii)
In the absence of disorder the azimuthal angular momentum is
conserved, which greatly reduces the computational difficulty of
the BdG diagonalization.

The order parameter field on the sphere is taken to be of the
following form \cite{Moller,ReadGreen}
\begin{eqnarray}   \label{Eq:Delta_vv}
    \Delta_{\rm v \bar{v}} & = & \Delta_p (\omgvec,\omgvec') F_{\rm v \bar{v}}(\bar{\omgvec}) \nonumber \\
        & = & \sum_{\begin{subarray}{c}lm\\l'm'\end{subarray}} \Delta_{lm,l'm'}
              Y_{-\half, l, m}(\omgvec)  Y_{-\half, l', m'}(\omgvec'), \nonumber \\
    \Delta_p (\omgvec,\omgvec')
        & = & \frac{\Delta_0} {(4\pi\xi_p^2) (l_F + \half)} \\
        &   & \times (\alpha \beta' - \beta \alpha')
                     | \alpha \alpha'^* + \beta \beta'^* |^{ 2(R/\xi_p)^2 },  \nonumber
\end{eqnarray}
which defines $\Delta_{lm,l'm'}$. $\Delta_0$ is the pairing
amplitude, the pairing range is $\xi_p$, and $l_F$ is the Fermi
angular momentum, given by $\epsilon_F = l_F (l_F+1)/(2m R^2)$. The
functions $\alpha = \cos(\theta/2)$ and $\beta = \sin(\theta/2)
\e^{-i\phi}$ are spinor functions. $Y_{q l m}$ are monopole
harmonics \cite{WuYang}, where $q,l,m$ are half odd integers. The
order parameter $\Delta_p (\omgvec,\omgvec')$ acquires a $2\pi$
phase when $\omgvec$ encircles $\omgvec'$, which describes \pxipy
pairing. $|\Delta_p|$ keeps the particles within the pairing range
$|\omgvec-\omgvec'| \sim \xi_p$.

The order parameter field $F_{\rm v \bar{v}}(\bar{\omgvec})$
describes the vorticity of the pair center of mass $\bar{\omgvec} =
(\omgvec + \omgvec')/2$. We choose $F_{\rm v \bar{v}}$ to describe
an antivortex on the north pole and a vortex on the south pole, with
the direction of vorticity defined relative to the chirality of the
\pxipy order parameter, depicted in Fig.~\ref{Fig:sphere}. For the
vortex pair field, we use the analytical form (without self
consistency) \cite{PS}
\begin{eqnarray}   \label{Eq:F_vv}
    F_{\rm v \bar{v}}(\omgvec)
        & = & \frac{\sin\theta \cdot R/\xi} {\sqrt{ 1 + (\sin\theta \cdot R/\xi)^2 } } \e^{i\phi} \nonumber\\
        & = & \sum_{L=1,3,5,\ldots } f_L Y_{L1}(\omgvec),
\end{eqnarray}
which defines $f_L$.  $Y_{LM}$ are spherical harmonics, and $\xi = {2 \epsilon_F/( \pi \Delta_0 k_F)}$ is Pippard's
coherence length. We take $\xi_p < \xi$ for simplicity.

The BdG equation is represented as a matrix in terms of 3j symbols as
\begin{eqnarray}   \label{Eq:BdG_matrix}
    T_{lm,l'm'} & = & \epsilon_F \frac{l(l+1)-{1 \over 4}} {l_F(l_F+1)-{1 \over 4}}
                      \, \delta_{ll'} \delta_{mm'},  \nonumber\\
    \Delta_{lm,l'm'} & = & \delta_{m', 1-m} \Delta_0  \sqrt{ (2l+1)(2l'+1) }  \nonumber \\
                     &   & \left( D_l + (-1)^{l+l'} D_{l'} \right)
                            \sum_{L} f_L \sqrt{ 2L+1 \over 16\pi }             \nonumber\\
                     &&~~~~~\times
                      \left( \begin{array}{ccc} l & l' & L \\ -m    & m-1    & 1 \end{array} \right)
                      \left( \begin{array}{ccc} l & l' & L \\ \half & -\half & 0 \end{array} \right)   \nonumber\\
    D_l & \simeq & \frac{l}{l_F} \e^{ (- l^2 + l_F ^2 )(\xi_p/R)^2 }. \nonumber
\end{eqnarray}

Diagonalizing Eq.~(\ref{Eq:BdG}) produces a set of energies $E_n$ and
corresponding eigenvectors $u_n^{lm} , v_n^{lm}$. By azimuthal symmetry,
$m$ is a good quantum number.
The BdG wavefunctions on the sphere are
\begin{eqnarray}   \label{Eq:un_vn}
    u_n(\omgvec) & = & \sum_{l} u_n^{lm} Y_{-\half,l,m}(\omgvec), \\
    v_n(\omgvec) & = & \sum_{l} v_n^{lm} Y_{-\half,l,-m+1}^*(\omgvec).
\end{eqnarray}
In Fig.~\ref{Fig:Em} we depict the BdG spectrum of the vortex pair
as a function of $m$. The continuum states above the gap
\mbox{$|E_n| > \Delta_0$} are extended, while the branch that
approaches zero is the \mbox{\pxipy} version of the CdGM core states.
Their number is of order $\epsilon_F/\Delta_0$, and their spacing
is of order $\delta_c$ \cite{KopninSalomaa}.

As seen in the inset of Fig.~\ref{Fig:Em}, each CdGM state is almost doubly
degenerate. The splitting represents weak tunneling between the north and
south pole core states. Indeed, we find that the tunnel splittings decrease
exponentially with the radius of the sphere
$\delta E_n \sim\e^{-R/\xi}$ for $R \gg \xi$.

The lowest positive energy approaches zero as $E_{0} \sim
\e^{-R/\xi}$. In the large sphere limit, the wavefunctions
$u_0(\Omega) \approx v_0(\Omega)$ are equally split between the
north and south poles. The corresponding BdG fermion is
constructed out of two well separated Majorana operators in the
north and south poles. We have also verified that $E_0$ is
insensitive to the addition of moderate potential disorder
\cite{TBP}. This agrees with previous asymptotic calculations in
the plane which have shown that the Majorana excitations are
'topologically protected' against perturbations \cite{Volovik,ReadGreen}.

The asymptotic predictions for the wavefunctions $u_0(r)$ for
core-less vortices in the plane \cite{GurarieRadzihosky} are
\begin{equation}   \label{Eq:u0}
    u_0({\bf x}) \sim \left\{
        \begin{array}{ll}
            J_0(k_Fr) \e^{-r/\pi\xi}             & \mbox{antivortex}, \\
            J_1(k_Fr) \e^{i \phi} \e^{-r/\pi\xi} & \mbox{vortex},
        \end{array}  \right.
\end{equation}
which are valid for {\em both} $r \gg \xi$ and $r \ll \xi$.
Numerically, we confirmed that these predictions hold even in finite
core sizes.

The physical reason behind the difference in Eq.~(\ref{Eq:u0}) is
that the Majorana wavefunctions are sensitive to the \emph{sum} of
vorticity and relative angular momentum. $J_0$ is obtained only when
that sum vanishes, and this is important for the experimental
signature we discuss below.


{\em Local Density of States (LDOS). }
At zero temperature the LDOS is defined as \cite{Shore}
\begin{equation}   \label{Eq:LDOS}
    {\cal T}(E,r) = \sum_{n} |u_n(r)|^2 \delta(E-E_n) + |v_n(r)|^2 \delta(E+E_n),
\end{equation}
where $r$ is the distance from the vortex (or antivortex) center.

\begin{figure}[htb] \vspace{-0.3cm}
\begin{center}
\includegraphics[width=9cm,angle=0]{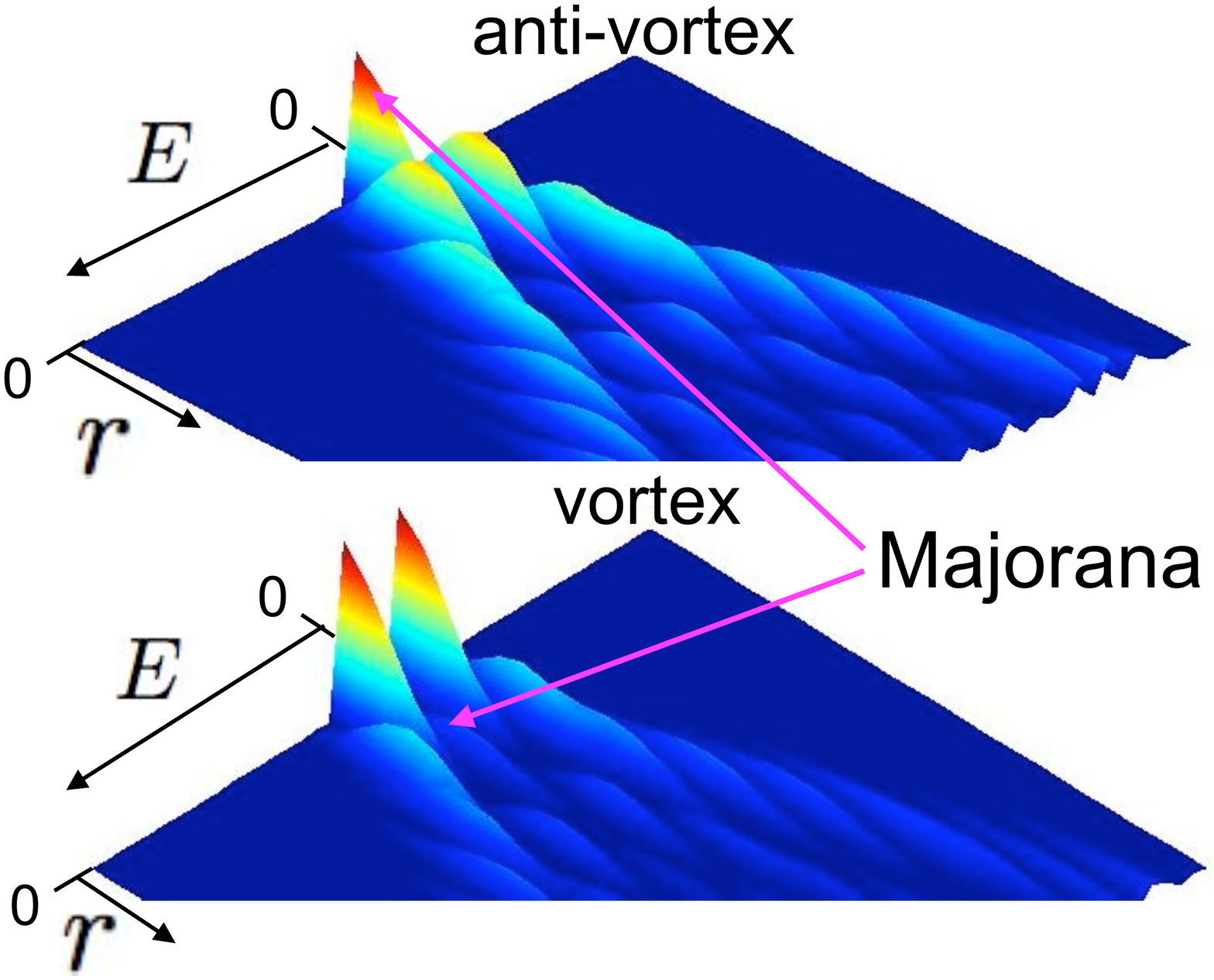}
\vspace{-0.8cm} %
\caption{   \label{Fig:LDOS} %
Zero temperature local density of states of Eq.~(\ref{Eq:LDOS}) near
the vortex and the antivortex centers. The peaks belong to the CdGM
states. Notice that the zero energy Majorana mode is removed from
the origin in the vortex, while it is maximized at the origin in the
antivortex. }
\end{center}
\end{figure}

Fig.~\ref{Fig:LDOS} shows the LDOS near the cores of the vortex
and the antivortex for displacements \mbox{$r \ll \xi$}, and
energies \mbox{$|E|\le \Delta_0$}. The Majorana state can be
easily discerned as the zero energy peak in both vortex and
antivortex cores.

The other CdGM core states also appear as oscillatory peaks, with
energy spacing $\delta_c$. The difference between the vortex and
antivortex excitations is apparent: In Eq.~\ref{Eq:u0} the
antivortex Majorana state is peaked at $r = 0$, while the other CdGM
states have nodes at $r = 0$. In contrast, the vortex Majorana state
is peaked at half a Fermi wavelength away from the origin, while two
CdGM states are peaked at $r = 0$. Notice that in the vortex, the
Majorana state has a significantly  lower peak than in the
antivortex.

In a tunneling spectroscopy experiment (e.g.~Ref.~\cite{Davis}),
the discrete LDOS spectrum is smeared by temperature broadening.
The tunneling conductance \cite{Gygi} is defined as
\begin{equation}  \label{Eq:dIdV}
    {d I\over dV}(E,r) \sim
    T \int dE' \left({ \partial f(E-E') \over \partial E' }\right) {\cal T}(E',r),
\end{equation}
where $f(E)$ is the Fermi-Dirac distribution at zero chemical
potential and temperature $T$.

In the BCS weak coupling  regime, $k_F \xi \gg 1$, and therefore
$\delta_c$  could be a very small temperature scale. At moderate
temperatures $\delta_c < T < \Delta_0$, the peaks of Fig.~\ref{Fig:LDOS}
are smeared on the energy axis (but not on the $r$ axis, therefore
an asymmetry effect can be observed.

\begin{figure}[htb]
\vspace{-0.3cm}
\begin{center}
\includegraphics[width=9cm,angle=0]{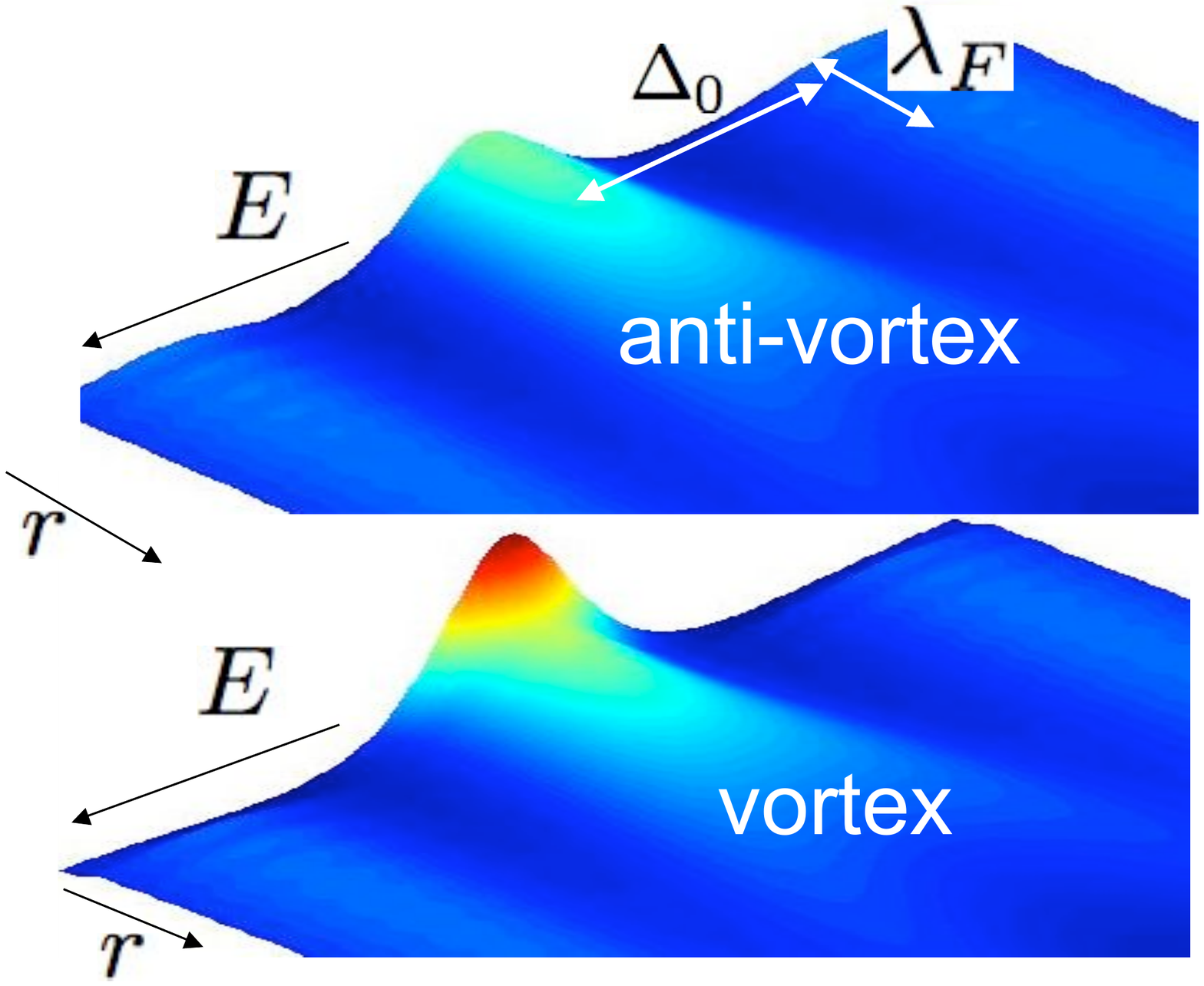}
\vspace{-0.8cm} %
\caption{    \label{Fig:dIdV} %
Tunneling conductance of Eq.~(\ref{Eq:dIdV}), in arbitrary units.
$\lambda_F$ is the fermi wavelength. The temperature \mbox{$T = 0.15
\Delta_0$}, is about 10 times larger than the CdGM level spacing.  }
\end{center}
\end{figure}

A typical tunneling conductance is depicted in Fig.~\ref{Fig:dIdV},
which shows a central peak at $r = 0, E = 0$, with low broad ridges
dispersing away to larger $r,E$. We see that \emph{the central peak
of the vortex is twice the height of that of the antivortex}.

This effect is a direct consequence of Eq.~(\ref{Eq:u0}). Under temperature
smearing the {\em two} CdGM peaks at $r = 0$ of the vortex, merge into
one large central peak. In contrast, only a  single Majorana state is
responsible for the central peak of the antivortex. Since the relevant
maximas in the LDOS are nearly identical, a ratio of 2 is obtained at
elevated temperatures.


Our effect requires having spatial resolution in tunneling conductance
better than a Fermi wavelength. If $dI/dV(r,E)$ is convoluted with areal
resolution of $(\delta r)^2 > \lambda_F^2$, the ratio between the vortex
and antivortex peak heights rapidly approaches unity as $\delta r  >
\lambda_F$. The ratio is weakly temperature dependent in the regime
$\delta_c < T < \Delta_0$.

In real three dimensional samples, zero bias peaks are somewhat
suppressed by bulk states and surface imperfections,
(Ref.~\cite{Davis} reports 15\% enhancement above the high voltage
background). Nevertheless, it is the ratio of 2 between the vortex
and an antivortex enhancement which would signal the Majorana state.
To avoid changes in the background, we suggest to leave the tip at
the same position while reversing the magnetic field. The field
should be localized and weak enough so as not to overturn the chiral
order parameter.

In real three dimensional samples, zero bias peaks are somewhat
suppressed by bulk states and surface imperfections,
(Ref.~\cite{Davis} reports a 15\% enhancement above the high voltage
background). Nevertheless, it is the {\em ratio of 2} between the
vortex and an antivortex enhancement which would signal the Majorana
state. To avoid changes in the background, we suggest to leave the
tip at the same position while reversing the magnetic field. The
field should be localized and weak enough so as not to overturn the
chiral order parameter.


{\em General Cases.}
Some difference between vortex and antivortex excitations is expected
for any chiral symmetry breaking (CSB) superconductor. The important
questions are whether this difference is observable at $T>\delta_c$,
and whether it is sensitive to the existence of Majorana fermions.

For a CSB superconductor with relative angular momentum $M = 1, 2,
\ldots $ (chiral-{\it p}, chiral-{\it d}, \ldots), there is a
Majorana state in the vortex core, provided the vorticity $N$ obeys
$N+M=0, \pm2, \pm 4, \ldots$ \cite{Volovik}. However, in most cases
these Majorana states vanish at the origin. The only exceptions are
cores of antivortices which satisfy $N = -M$. Our factor of 2 effect
will be observable only for this subset of cases. Notice that \pxipy
is the only case where the effect occurs for vorticity $|N|=1$.

Since the core states are only sensitive to large potential
gradients, moderate disorder does not destroy the Majorana states.
We have explicitly confirmed this expectation numerically
\cite{TBP}, by solving the BdG equation with a white noise
potential. For disorder potential fluctuations up to order
$\epsilon_F$, the Majorana tunneling energy decays with $R$, and
$u_0(r)$ with $r$, with the same exponents as the clean system, and
the peak height doubling signature of the Majorana states is
essentially unaffected.


{\em Summary. } We solved the BdG spectrum of \mbox{\pxipy} vortex
pair state in the spherical geometry. We show that even at high
temperatures compared to the CdGM state spacing, a signature of the
Majorana state remains when one compares the LDOS of the vortex to
that of the antivortex.


{\em Acknowledgements.} We thank Ady Stern for useful discussions.
Support from US - Israel Binational Science foundation and Israel
Science Foundation is acknowledged. AA acknowledges Aspen Center for
Physics for its hospitality. HAF acknowledges the support of the NSF
through Grant No. DMR-0704033. AA and SHS acknowledge the
hospitality of the KITP where this collaboration was initiated.



\begin{thebibliography}{99}

\bibitem{NayakRMP}
    S.~Das Sarma, M.~Freedman, C.~Nayak, S.~Simon and A.~Stern,
        Rev.~Mod.~Phys.~\textbf{80}, 1083 (2008), and references therein.

\bibitem{Bravyi}
    While such systems are not universal for topological quantum
    computation, they could serve as quantum memories. Schemes
    have also been constructed for partially topological quantum
    computation. See
    S.~Bravyi, Phys.~Rev.~A \textbf{73}, 042313 (2006).

\bibitem{He3Volovik}
    G.E.~Volovik, The Universe in a Helium Droplet (Clarendon Press, Oxford, 2003).

\bibitem{RiceSigrist} 
    T.M.~Rice and M.~Sigrist, J.~Phys.~Cond.~Matter \textbf{7}, L643 (1995);
    G.~Baskaran, Physica B \textbf{223\&224}, 490 (1996);
    Y.~ Maeno. T.M.~Rice and M.~Sigrist, Phys.~Today 54, No.~1, 42 (2001).

\bibitem{MooreRead}
    G.~Moore and N.~Read, Nucl.~Phys.~B \textbf{360}, 362 (1991).

\bibitem{Grieter}
    M.~Greiter, X.G.~Wen and F.~Wilczek, Nucl.~Phys.~B \textbf{374}, 567 (1992).

\bibitem{GurarieRadzihosky} 
    V.~Gurarie and L.~Radzihovsky, Annals of Physics \textbf{322}, 2 (2007).

\bibitem{ReadGreen} 
    N.~Read and D.~Green, Phys.~Rev.~B \textbf{61}, 10267 (2000).

\bibitem{Volovik} 
    G.E.~Volovik, Pis'ma Zh.~Eksp.~Teor.~Fiz.~\textbf{70}, 601 (1999)
        [JETP Lett.~\textbf{70}, 609 (1999)].

\bibitem{TewariKim}
   In \SRO $\,$ the ``full-quantum vortex'' (a vortex in the order
   parameter of both spin components) is lower energy than the half-quantum
   vortex, so that the half-quantum vortex does not naturally occur.
   Nonetheless, several proposals have appeared for how to stabilize
   the half quantum vortices. See
   S.~Das Sarma, C.~Nayak and S.~Tewari, Phys.~Rev.~B \textbf{73}, 220502 (2006);
   S.B.~Chung, H.~Bluhm, and E.-A.~Kim Phys.~Rev.~Lett.~\textbf{99}, 197002 (2007);
   In \He3A, see:
   M.M.~Salomaa and G.E.~Volovik, Phys.~Rev.~Lett.~\textbf{55}, 1184 (1985).

\bibitem{e4quantumHall}
    For the 5/2 state there are now several experiments that claim to
    observe the e/4 quasiparticle, which would be the Majorana-fermion-carrying
    vortex.
    See M.~Dolev \etal Nature \textbf{452}, 829 (2008);
    I.P.~Radu \etal Science \textbf{320}, 899 (2008);
    R.L.~Willett, M.J.~Manfra, L.~N.~Pfeiffer and K.W.~West, arXiv:cond-mat/0807.0221.

\bibitem{DemlerTewari}
    Proposals based on tunneling of the Majorana between two vortices were given by
    C.J.~Bolech and E.~Demler, Phys.~Rev.~Lett.~\textbf{98}, 237002 (2007);
    S.~Tewari, C.~Zhang, S.~Das Sarma, C.~Nayak and D.H.~Lee,
        Phys.~Rev.~Lett.~\textbf{100}, 027001 (2008).

\bibitem{Shore} 
    J.D.~Shore, M.~Huang, A.T.~Dorsey and J.Sentha,
        Phys.~Rev.~Lett.~\textbf{62}, 3089 (1989).

\bibitem{endnote1}
    For \He3A one can at least imagine tunneling an atom through a
    nanoconstriction at varying pressure although in practice this
    might be extremely difficult. For the 5/2 state, tunneling in or
    out of the system requires flux-attachment (see Ref.~\cite{ReadGreen})
    and would likely result in a strong pseudogap in the tunneling
    amplitude. See for example,
    S.~He, P.~M.~Platzman and B.~I.~Halperin, Phys.~Rev.~Lett.~\textbf{71}, 777 (1993).

\bibitem{CdGM}
    C.~Caroli, P.G.~de Gennes and J.~Matricon, Phys.~Lett.~\textbf{9}, 307 (1964).

\bibitem{KopninSalomaa} 
    N.B.~Kopnin and M.M.~Salomaa, Phys.~Rev.~B \textbf{44}, 9667 (1991).

\bibitem{BdG}
    P.G.~de Gennes, Superconductivity of Metals and Alloys
        (WA Benjamin Inc., New York, 1966).

\bibitem{Moller}
    J.K.~Jain and R.K.~Kamilla, Phys.~Rev.~B \textbf{55}, R4895 (1997);
    G.~Moller and S.~H.~Simon, Phys.~Rev.~B \textbf{77}, 075319 (2008).

\bibitem{WuYang} 
    T.T.~Wu and C.N.~Yang, Nucl.~Phys.~B \textbf{107}, 365 (1976);
                           Phys.~Rev.~D \textbf{16}, 1018 (1977).

\bibitem{PS}
    C.~Pethik and H.~Smith, Bose Einstein Condensation in Dilute Gases
        (Cambridge University press, Cambridge, 2002).

\bibitem{TBP}
    Y.E.~Kraus and A.~Auerbach, to be published.

\bibitem{Davis} 
    C.~Lupien, S.K.~Dutta, B.I.~Barker, Y.~Maeno and J.C.~Davis, arXiv:cond-mat/0503317.

\bibitem{Gygi}
    F.~Gygi and M.~Schl\"{u}ter, Phys.~Rev.~B \textbf{41}, 822 (1990);
                                 Phys.~Rev.~B \textbf{43}, 7609 (1991).

\end{thebibliography}
\end{document}